\documentclass[envcountsame, runningheads,orivec]{llncs} %
\AtBeginDocument{\DeclareRobustCommand{\doi}{\begingroup\catcode`\_=12 \doiaux}}
\newcommand*{\doiaux}[1]{\href{https://doi.org/#1}{doi: #1}\endgroup}
\usepackage{hyperref}
\usepackage{amsmath}
\usepackage{amssymb}
\usepackage{paralist}
\usepackage{relsize}
\usepackage{algorithm}
\usepackage{algorithmic}
\usepackage{stmaryrd}
\usepackage{tikz}%
\usepackage{bussproofs}
\usepackage{bm}

\usetikzlibrary{shapes,arrows,chains,positioning,
decorations,automata,backgrounds,petri,matrix,fit,decorations.markings,decorations.pathreplacing,calc}

\tikzset{
  every edge/.append style = {
    >=latex,
  line width = 1pt, ->, auto
  }, 
}

\def\symboleq{\mathbin{\bm{=}}}
\def\symbollor{\mathbin{\bm{\lor}}}
\def\symbolto{\mathbin{\bm{\to}}}
\def\symbolleftrightarrow{\mathbin{\bm{\leftrightarrow}}}
\def\symbolland{\mathbin{\bm{\land}}}
\def\symbolexists{\bm{\exists}}
\def\symbolforall{\bm{\forall}}
\def\symbollnot{\bm{\lnot}}

\def\relcdot{\boldsymbol{\cdot}}

\def\jump#1{\llbracket#1\rrbracket}

\def\dom{\operatorname{dom}}
\def\cod{\operatorname{cod}}

\def\Term{\mathcal{T}}

\def\FO{\mathsf{FO}}

\def\classcoNEXPTIME{\mathsf{coNEXPTIME}}

\def\defiff{\;:\Longleftrightarrow\;}

\def\bigland{\bigwedge}

\def\α{\alpha}
\def\β{\beta}
\def\γ{\gamma}\def\Γ{\Gamma}
\def\δ{\delta}\def\Δ{\Delta}
\def\ε{\epsilon}
\def\φ{\varphi}
\def\ψ{\psi}
\def\τ{\tau}\def\Τ{T}
\def\ρ{\rho}
\def\μ{\mu}
\def\λ{\lambda}\def\Λ{\Lambda}
\def\σ{\sigma}\def\Σ{\Sigma}
\def\θ{\theta}\def\Θ{\Theta}
\def\π{\pi}\def\Π{\Pi}
\def\ω{\omega}
\def\χ{\chi}\def\Χ{X}
\def\×{\times}
\def\∪{\cup}\def\∩{\cap}\def\⊆{\subseteq} \def\∈{\in}
\def\∧{\land}\def\∨{\lor}\def\→{\to}

\def\∀{\forall}\def\∃{\exists}
\def\⇔{\iff}\def\⇒{\Rightarrow}\def\⇐{\Leftarrow}
\def\≡{\equiv}
\def\<{\langle}\def\>{\rangle}
\def\∵{\because}\def\∴{\therefore}
\def\bigland{\bigwedge}

\def\assignment{\mathbin{:=}}

\def\@{\mathsf{@}}

\def\converse{ {\mbox{\protect\scalebox{.5}{$\smallsmile$}}} }

\def\CF{\mathrm{CF}}
\def\IF{\mathrm{IF}}

\def\model{M} 
\usepackage[utf8]{inputenc} %
\usepackage[T1]{fontenc} %
\usepackage{mathpazo} %
\usepackage[autostyle=true]{csquotes} %

\spnewtheorem*{remark*}{Remark}{\itshape}{\rmfamily}
\begin{document}

\title{The Undecidability of FO3 and the Calculus of Relations with Just One Binary Relation\thanks{Supported by JSPS KAKENHI Grant Number 16J08119.
The affiliation above is that at the time of the conference version. The e-mail address is the author's address at the time of uploading to the repository.
}\fnmsep\thanks{This is a long version of the paper at ICLA'19 \cite{nakamuraUndecidabilityFO3Calculus2019}, with (Ax.1) and (Ax.2) in Sect.~\ref{section: Reductions to the One Binary Relation Case} corrected (see the remark at the end of Sect.~\ref{section: Reductions to the One Binary Relation Case}).}}
\titlerunning{The Undecidability of FO3 with Just One Binary Relation}
\author{Yoshiki Nakamura}
\authorrunning{Yoshiki Nakamura}
\institute{
	Tokyo Institute of Technology, Japan\\
	\email{nakamura.yoshiki.ny@gmail.com}
}
\maketitle

\begin{abstract}
	The validity problem for first-order logic is a well-known undecidable problem.
  The undecidability also holds even for $\mathsf{FO}3$ and (equational formulas of) the calculus of relations.
	In this paper we tighten these undecidability results to the following:
	(1) $\mathsf{FO}3$ with just one binary relation is undecidable even without equality; and
	(2) the calculus of relations with just one character and with only composition, union, and complement is undecidable.
  Additionally we prove that the finite validity problem is also undecidable for the above two classes.
  \keywords{First-order logic \and The calculus of relations \and Undecidability}
\end{abstract}

\section{Introduction}
The \emph{validity problem} for first-order logic (a.k.a. the Entscheidungsproblem) is the problem to decide whether a given formula of first-order logic ($\FO$) is valid or not.
The problem is recursively enumerable by G{\"o}del's completeness theorem \cite{godelVollstandigkeitAxiomeLogischen1930},
but is undecidable, as shown by Church \cite{Church36} and Turing \cite{Turing37}.
In connection with the undecidability result, many decidable and undecidable variants of $\FO$ have been studied (we refer to the book \cite{borgerClassicalDecisionProblem1997}).

One restriction is to consider the validity problem over a restricted signature.
In 1915, L{\"o}wenheim \cite{lowenheimUberMoglichkeitenIm1915}
proved that monadic $\FO$ with equality is decidable (more precisely, $\classcoNEXPTIME$-complete \cite[p.~318]{lewisComplexityResultsClasses1980}).
Monadic $\FO$ is $\FO$ with only unary relation symbols.
In 1919, Skolem \cite{skolemUntersuchungenUberAxiome1919} extended the decidability result to monadic second-order logic (see also \cite{vardiPhilosophicalIndustrialLogics2008}).
Subsequently B{\"u}chi \cite{Buchi1962} extended the decidability result to monadic second-order logic over $\mathbb{N}$ with the linear order (called $\mathsf{S1S}$).
In contrast to this, $\FO$ with just one binary relation symbol is undecidable even without equality.
L{\"o}wenheim \cite{lowenheimUberMoglichkeitenIm1915} reduced $\FO$ to $\FO$ with only binary relation symbols, so the latter is also undecidable by \cite{Church36,Turing37}, but the number of binary relation symbols is countably infinite.
Subsequently the number was reduced to three by Herbrand \cite{herbrandProblemeFondamentalLogique1931} and to just one by Kalm{\'a}r \cite{kalmarZuriickftihrungEntscheidungsproblemsAuf1936}, see e.g., \cite[p.~6]{borgerClassicalDecisionProblem1997} or \cite[Theorem 21.4 (The Church-Herbrand theorem)]{boolosComputabilityLogic2007}.
Over finite models, $\FO$ with just one binary relation symbol is also undecidable by Trakhtenbrot's Theorem \cite{trakhtenbrotImpossibilityAlgorithmDecision1963} (moreover, the undecidability holds even over finite graphs of vertex-degree at most $3$ \cite{willardHereditaryUndecidabilityTheories1994}).

Another restriction is with respect to the number of variables occurring in formulas.
$\FO k$ denotes the restriction of $\FO$ to formulas with at most $k$ distinct variables.
In these classes, it is known that $\FO 3$ with countably infinitely many binary relation symbols and without equality is undecidable (because the $\symbolforall\symbolexists\symbolforall$ case is undecidable \cite{Kahr62}),
whereas $\FO 2$ (even with equality) is decidable \cite{mortimerLanguagesTwoVariables1975} and $\classcoNEXPTIME$-complete \cite[Corollary 5.4]{gradelDecisionProblemTwoVariable1997}.
$\FO 2$ has a connection with modal logic in the sense that some propositional modal logics can be embedded into $\FO 2$ {\cite{vardiWhyModalLogic1997}}.
With respect to the expressive power of unary relations, $\FO 2$ is equivalent to Boolean modal logic with relational converse and the identity relation {\cite{lutzModalLogicTwoVariable2001}}.
As for $\FO 3$, it has a connection with \emph{the calculus of (binary) relations} \cite{tarskiCalculusRelations1941}.
We denote the full signature of the calculus of relations by $\<\cdot, \bullet^-, \cup, \bullet^{\converse}, 1\>$, where
$\cdot$ is relational composition,
$\bullet^-$ is set-theoretic complement,
$\cup$ is set-theoretic union,
$\bullet^{\converse}$ is relational converse, and
$1$ is the identity relation.
Actually, with respect to the expressive power of binary relations, $\FO 3$ with equality is equivalent to terms of the calculus of relations {\cite{tarskiFormalizationSetTheory1987}} (see also \cite{givantCalculusRelationsFoundation2007}).

In this paper we prove that the intersection of the above two restrictions is also undecidable, i.e., the validity problem for $\FO 3$ with just one binary relation symbol and without equality is undecidable.
Moreover we prove that the finite validity problem for the class is undecidable.
The class is the minimal undecidable fragment with respect to the above two restrictions since both monadic $\FO$ and $\FO 2$ are decidable.
As for the calculus of relations,
we prove that the validity problem and the finite validity problem
for equational formulas of the calculus of relations with just one character over the signature $\<\cdot, \bullet^{-}, \cup\>$ are undecidable.

\subsection*{Outline}
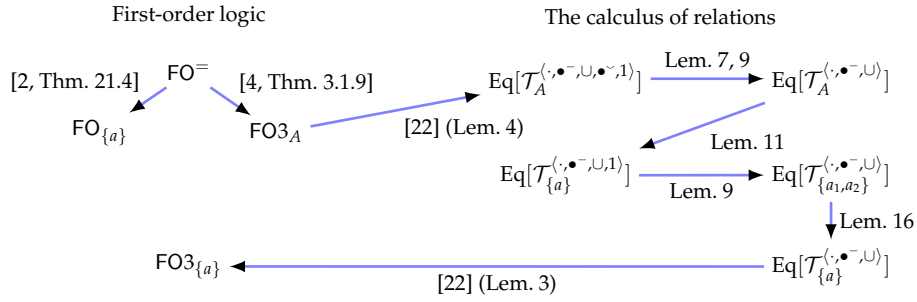
\begin{figure}[ht]
	\centering
  \begin{tikzpicture}[every node/.style={scale=.9}]
    \node(FO_label){First-order logic};
		\node(FO)[below = .3cm of FO_label]{$\FO^=$};
		\node(FOsep)[below = .5cm of FO]{};
		\node(FOch)[left = .6cm of FOsep]{$\FO_{\{a\}}$};
		\node(FO3)[right = .6cm of FOsep]{$\FO3_{A}$};
    \node(REL_label)[right = 3.5cm of FO_label]{The calculus of relations};
		\node(REL)[below left = .3cm and -1.5cm of REL_label]{$\mathrm{Eq}[\mathcal{T}_{A}^{\<\cdot, \bullet^-, \cup, \bullet^{\converse}, 1\>}]$};
		\node(RELc1)[right = 1.5cm of REL]{$\mathrm{Eq}[\mathcal{T}_{A}^{\<\cdot, \bullet^-, \cup\>}]$};

    \node(RELc_1)[below = .6cm of REL]{$\mathrm{Eq}[\mathcal{T}_{\{a\}}^{\<\cdot, \bullet^-, \cup, 1\>}]$};
    \node(RELc1_2)[below = .6cm of RELc1]{$\mathrm{Eq}[\mathcal{T}_{\{a_1,a_2\}}^{\<\cdot, \bullet^-, \cup\>}]$};
    \node(RELc1_1)[below = .5cm of RELc1_2]{$\mathrm{Eq}[\mathcal{T}_{\{a\}}^{\<\cdot, \bullet^-, \cup\>}]$};
		\node(FO3_1)[below = 1.4cm of FOsep]{$\FO3_{\{a\}}$};

		\path
		(FO) edge[draw = blue!50] node{{\cite[Thm. 3.1.9]{borgerClassicalDecisionProblem1997}}} (FO3)
		(FO) edge[draw = blue!50, above left] node{{\cite[Thm. 21.4]{boolosComputabilityLogic2007}}} (FOch)
		(FO3) edge[draw = blue!50, below right] node{{\cite{tarskiFormalizationSetTheory1987}} (Lem. \ref{lemma: FO3 to rel})} (REL)
		(REL) edge[draw = blue!50, align = center] node{{Lem. \ref{lemma: converse elimination 2}}, {\ref{lemma: identity elimination}}} (RELc1)
		(RELc1) edge[draw = blue!50] node{Lem. \ref{lemma: to one using identity}} (RELc_1)
		(RELc_1) edge[draw = blue!50, below] node{Lem. \ref{lemma: identity elimination}} (RELc1_2)
		(RELc1_2) edge[draw = blue!50] node{Lem. \ref{lemma: to one with none}} (RELc1_1)
		(RELc1_1) edge[draw = blue!50, below] node{{\cite{tarskiFormalizationSetTheory1987}} (Lem. \ref{lemma: rel to FO3})} (FO3_1);

	\end{tikzpicture}
	\caption{An overview of reductions}\label{figure : translation}
\end{figure}

Figure~\ref{figure : translation} gives the outline of this paper.
Every arrow denotes a reduction from the source to the target preserving validity and finite validity
(hence a \emph{conservative reduction}, via taking the negations of the source and target formulas),
and $A$ denotes a countably infinite set.
In Sect.~\ref{section: FO3 and the Calculus of Relations}
we introduce first-order logic and the calculus of relations,
and we show that $\FO 3$ with equality and the calculus of relations are equivalent in the sense of expressive power of binary relations.
In this section we prove that relational converse $\bullet^{\converse}$ and the identity relation symbol $1$ can be eliminated by using fresh characters
preserving validity and finite validity.
In Sect.~\ref{section: Reductions to the One Binary Relation Case}
we give reductions for reducing the number of characters to one.
Finally, in Sect.~\ref{section: conclusion} we conclude this paper.

\section{$\FO 3$ and the Calculus of Relations}\label{section: FO3 and the Calculus of Relations}
Let $V$ be a countably infinite set of \emph{(first-order) variables}.
In this paper we assume that every \emph{signature} of first-order logic is a set of binary relation symbols.
Let $A$ be a set denoting a signature.
The set of \emph{formulas} of first-order logic over $A$, written $\FO^=_{A}$, is defined by the following grammar 
(we omit parentheses and use them only when it is not clear how to parse):
{\small
  \begin{align*}
    \varphi, \psi \in \FO^=_{A} & ::= a(x,y) \mid x \symboleq y \mid \varphi \symbollor \psi \mid \symbolexists x. \varphi \mid \symbollnot \varphi 
                              \tag*{$a \in A$ and $x,y \in V$}
  \end{align*}
}
We use the following abbreviations:
\begin{inparaenum}
	\item $\varphi \symbolland \psi :\equiv \symbollnot ((\symbollnot \varphi) \symbollor (\symbollnot \psi))$;
	\item $\varphi \symbolto \psi :\equiv (\symbollnot \varphi) \symbollor \psi$;
	\item $\varphi \symbolleftrightarrow \psi :\equiv (\varphi \symbolto \psi) \symbolland (\psi \symbolto \varphi)$;
	\item $\symbolforall x. \varphi :\equiv \symbollnot \symbolexists x. \symbollnot \varphi$.
\end{inparaenum}
$\FO_{A}$ denotes the subset of $\FO^=_{A}$ whose formulas do not contain the symbol $=$.
$\FO k^=_{A}$ (resp. $\FO k_{A}$) denotes the subset of $\FO^=_{A}$ (resp. $\FO_{A}$) whose formulas contain at most $k$ distinct variables\footnote{For example, $\symbolexists x.\symbolexists y.\symbolexists z. \symbolexists w. a(x,y)$ is not an $\FO 3_{\{a\}}$ formula (the formula is an $\FO 4_{\{a\}}$ formula),
	but $\symbolexists x.\symbolexists y.\symbolexists z. \symbolexists x. a(x,y)$ is an $\FO 3_{\{a\}}$ formula.}.
$\FO^=$ denotes the set of all formulas in $\FO^=_{A}$ over a signature $A$.

A \emph{model} over $A$ is a tuple $\model = \<|\model|, \mathcal{R}_{\model}\>$,
where $|\model|$ is a nonempty set; and $\mathcal{R}_{\model}\colon A \to \wp(|\model|^2)$ is a function.
Each $\mathcal{R}_{\model}(a)$ denotes a binary relation on $|\model|$.
We may view every model as a graph, e.g.,
each element of $|\model|$ is called a \emph{vertex} and each element of $\mathcal{R}_{\model}(a)$ is called an \emph{edge} labeled with $a$.
An \emph{interpretation} $I$ on $\model$ is a function from $V$ to $|\model|$.
The \emph{semantics} of every formula $\varphi$ on $\<\model,I\>$ is a truth value.
$\<\model,I\> \models \varphi$ (resp. $\<\model,I\> \not\models \varphi$) denotes that the value of $\varphi$ on $\<\model,I\>$ is $\mathit{true}$ (resp. $\mathit{false}$).
$\<\model,I\> \models \varphi$ is defined as follows:
\begin{inparaenum}
	\item $\<\model,I\> \models a(x,y) \defiff \<I(x),I(y)\> \in \mathcal{R}_{\model}(a)$ for $a \in A$;
	\item $\<\model,I\> \models x \symboleq y \defiff I(x) = I(y)$;
	\item $\<\model,I\> \models \varphi \symbollor \psi \defiff \<\model,I\> \models \varphi \text{ or } \<\model,I\> \models \psi$;
	\item $\<\model,I\> \models \symbolexists x.\varphi$ $\defiff$  there is $v \in |\model|$ s.t. $\<\model,[v/x]I\> \models \varphi$;
	\item $\<\model,I\> \models \symbollnot \varphi \defiff \<\model,I\> \not \models \varphi$,
\end{inparaenum}
where $[v/x]I$ denotes the function $I$ in which $I(x)$ has been replaced by the element $v$.
Then the \emph{semantics} of $\varphi$ on $\model$ is defined by $\jump{\varphi}_{\model} \assignment \{I \colon V \to |\model| \mid \<\model,I\> \models \varphi \}$.
$\model \models \varphi$ denotes that, for any interpretation $I$, $\<\model,I\> \models \varphi$ holds.
$\jump{\varphi}_{\model}^{\vec{x}}$ denotes the \emph{projection} of $\jump{\varphi}_{\model}$ with respect to $\vec{x}$, i.e.,
$\jump{\varphi}_{\model}^{\vec{x}} \assignment \{\<I(x_1), \dots, I(x_n)\> \mid I \in \jump{\varphi}_{\model}\}$,
where $\vec{x} = \<x_1, \dots, x_n\>$.

A formula $\varphi$ over the signature $A$ is called \emph{valid} (resp. \emph{finitely valid}) if, for any model (resp. finite model) $\model$ over $A$,
$\model \models \varphi$ holds.
The \emph{validity problem} (resp. the \emph{finite validity problem}) is the problem to decide whether a given formula is valid (resp. finitely valid).
In this paper we rely on the undecidability of the $\symbolforall\symbolexists\symbolforall$ case {\cite{Kahr62}} and modify the result for $\FO 3$ and the calculus of relations.
This class is a \emph{conservative reduction} class \cite[Definition 2.1.35]{borgerClassicalDecisionProblem1997}
(i.e., there is a recursive function from every first-order logic formula over any signature to a formula of the class preserving satisfiability and finite satisfiability).
From this, the following holds.
\begin{lemma}[A corollary of {\cite[Theorem 3.1.9]{borgerClassicalDecisionProblem1997}}]\label{lemma: FO3 is undecidable}
  Let $A$ be any countably infinite set.
  Then
  \begin{itemize}
		\item the validity problem for $\FO 3_{A}$ is undecidable; and
		\item the finite validity problem for $\FO 3_{A}$ is undecidable.
	\end{itemize}
\end{lemma}

We now define the calculus of relations.
Let $X$ be a set and let $R$ and $R'$ be binary relations on $X$.
The \emph{identity relation} $\triangle(Y)$ on a set $Y$ is defined as $\{\<v,v\> \mid v \in Y\}$.
The \emph{relational composition} $R \relcdot R'$ is defined as
$\{\<v,v'\> \mid \exists v''.\<v,v''\> \in R \land \<v'',v'\> \in R'\}$.
The \emph{relational converse} $R^{\converse}$ is defined as $\{\<v',v\> \mid \<v,v'\> \in R\}$.
The \emph{union} $R \cup R'$ and the \emph{complement} $R^{-}$ are defined
as set-theoretic union and complement (in particular, $R^{-} \assignment X^2 \setminus R$). 
Let $A$ be a set of \emph{characters}.
The set of \emph{terms} over $A$, written $\Term_A$, is defined by the following grammar:
\begin{align*}
	t, u \in \Term_A & ::= a \mid t \cdot u \mid t^{-} \mid t \cup u \mid t^{\converse} \mid 1 \tag*{$a \in A$}
\end{align*}
We use the following abbreviations:
\begin{inparaenum}
  \item $t \cap u :\equiv (t^- \cup u^-)^-$;
  \item $\top :\equiv a \cup a^-$;
  \item $0 :\equiv a \cap a^-$;
  \item $t^{n+1} :\equiv t^n \cdot t$ and $t^1 :\equiv t$,
\end{inparaenum}
where $a$ is some character in $A$ and $n \ge 1$.
$\Term_A^{\varsigma}$ denotes the set of terms over $A$ and over the signature $\varsigma$, where $\varsigma$ is a subsequence of $\<\cdot, \bullet^-, \cup, \bullet^{\converse}, 1\>$.
Note that the symbols, $\bullet^{\converse}$ and $1$, are not used in these abbreviations.

Let $\model$ be a model over $A$.
The semantics of every term $t$ on $\model$, written $\jump{t}_{\model}$, is a binary relation on $|\model|$, defined as follows:
\begin{inparaenum}
	\item $\jump{a}_{\model} \assignment \mathcal{R}_{\model}(a)$;
	\item $\jump{t \cdot u}_{\model} \assignment \jump{t}_{\model} \relcdot \jump{u}_{\model}$;
	\item $\jump{t^{-}}_{\model} \assignment |\model|^2 \setminus \jump{t}_{\model}$;
	\item $\jump{t \cup u}_{\model} \assignment \jump{t}_{\model} \cup \jump{u}_{\model}$;
	\item $\jump{t^{\converse}}_{\model} \assignment \jump{t}_{\model}^{\converse}$;
	\item $\jump{1}_{\model} \assignment \triangle(|\model|)$.
\end{inparaenum}
Note that $\jump{t \cap u}_{\model} = \jump{t}_{\model} \cap \jump{u}_{\model}$,
$\jump{\top}_{\model} = |\model|^2$, and $\jump{0}_{\model} = \emptyset$ hold.
We also define the set of \emph{formulas of the calculus of relations} over a set $\Term$ of terms, written $\Phi[\Term]$, by the following grammar:
\begin{align*}
  \varphi,\psi \in \Phi[\Term] & ::= t \symboleq u \mid \varphi \symbollor \psi \mid \symbollnot \varphi
   \tag*{$t, u \in \Term$}
\end{align*}
The semantics of every formula $\varphi \in \Phi[\Term]$ on $\model$ is a truth value.
$\model \models \varphi$ (resp. $\model \not\models \varphi$) denotes that the value of $\varphi$ on $\model$ is $\mathit{true}$ (resp. $\mathit{false}$) defined as follows:
\begin{inparaenum}
  \item $\model \models t \symboleq u \defiff \jump{t}_{\model} = \jump{u}_{\model}$;
  \item $\model \models \varphi \symbollor \psi \defiff \model \models \varphi \text{ or } \model \models \psi$;
  \item $\model \models \symbollnot \varphi \defiff \model \not\models \varphi$.  
\end{inparaenum}
A formula of the form $t \symboleq u$ is called an \emph{equational formula}.
We denote the set of \emph{equational formulas} over a set $\Term$ by $\mathrm{Eq}[\Term]$.
Actually every formula is equivalent to an equational formula by the following lemma.
\begin{lemma}[cf. {\cite[Theorem 1]{madduxCalculusRelations2006}}]\label{lemma: equational formula}
  For any formula $\varphi$ in $\Phi[\Term_{A}]$, there is an equational formula $\psi$ of the form $t \symboleq 0$ in $\mathrm{Eq}[\Term_{A}]$ s.t. 
  for any $\model$ over $A$, $\model \models \varphi \iff \model \models \psi$.
\end{lemma}
\begin{proof}[Sketch]
  It is proved by using the following equivalences:
  \begin{inparaenum}
    \item $\model \models t \symboleq u
      \iff \model \models (t \cap u^-) \cup (t^- \cap u) \symboleq 0$;
    \item $\model \models (t \symboleq 0) \symbollor (u \symboleq 0)
      \iff \model \models t \cdot \top \cdot u \symboleq 0$;
    \item $\model \models \lnot (t \symboleq 0)
      \iff \model \models (\top \cdot t \cdot \top)^- \symboleq 0$.
  \end{inparaenum}
\end{proof}
By Lemma~\ref{lemma: equational formula}, every formula of the calculus of relations can be translated to an equational formula preserving validity and finite validity.
For simplicity, we may use formulas of the calculus of relations (but it is not essential).
Note that the symbols, $\bullet^{\converse}$ and $1$, are not used in the translation in Lemma~\ref{lemma: equational formula}.

We use $[n]$ to denote the set $\{1,\dots, n\}$ for every natural number $n \ge 0$.
The following lemma shows that every term of the calculus of relations has an equivalent $\FO 3^=$ formula.
\begin{lemma}[e.g., {\cite[p.~28]{tarskiFormalizationSetTheory1987}}, {\cite{givantCalculusRelationsFoundation2007}}]\label{lemma: rel to FO3}
	Let $x_1$, $x_2$, and $x_3$ be three distinct variables.
  There is a recursive function $G$ from $\Term_A \times [3]^2$ to $\FO 3^=_{A}$ s.t.,
  for any $\<t,i,j\>$ with $i \neq j$ and any model $\model$ over $A$, 
	$\jump{t}_{\model} = \jump{G(t,i,j)}_{\model}^{x_i,x_j}$ holds.
\end{lemma}
\begin{proof}[Sketch]
	We define $G(t, i, j)$ as follows:
	\begin{inparaenum}
		\item $G(a, i, j) \assignment a(x_i, x_j)$;
		\item $G(t \cdot u, i, j) \assignment \symbolexists x_k. G(t, i, k) \symbolland G(u, k, j)$;
		\item $G(t^-, i, j) \assignment \symbollnot G(t, i, j)$;
		\item $G(t \cup u, i, j) \assignment G(t, i, j) \symbollor G(u, i, j)$;
		\item $G(t^{\converse}, i, j) \assignment G(t, j, i)$;
		\item $G(1, i, j) \assignment x_i \symboleq x_j$,
	\end{inparaenum}
	where $k$ is the minimum element in $[3] \setminus \{i, j\}$.
	Then $\jump{t}_{\model} = \jump{G(t,i,j)}_{\model}^{x_i, x_j}$ is proved by induction on the structure of $t$.
\end{proof}
Actually the converse of Lemma~\ref{lemma: rel to FO3} also holds, i.e., $\FO 3^=$ and the calculus of relations are equivalent in the sense of expressive power of binary relations.
\begin{lemma}[e.g., {\cite[Sect. 3.9]{tarskiFormalizationSetTheory1987}}, {\cite{givantCalculusRelationsFoundation2007}}]\label{lemma: FO3 to rel}
	There is a recursive function $H$ from $\FO3^=_{A} \times V^2$ to $\Term_A$ s.t.,
  for any $\<\varphi, x, y\>$ and any model $\model$ over $A$,
  $\jump{\varphi}_{\model}^{x,y} = \jump{H(\varphi,x,y)}_{\model}$.
\end{lemma}
By Lemmas~\ref{lemma: rel to FO3} and \ref{lemma: FO3 to rel},
every equational formula of the calculus of relations can be translated to an $\FO 3$ sentence, and vice versa, by the following:
\begin{align*}
  \model \models t \symboleq u &\iff \model \models \symbolforall x_1.\symbolforall x_2. G(t, 1, 2) \symbolleftrightarrow G(u, 1, 2)\\
  \model \models \varphi &\iff \model \models H(\varphi, x, y) \symboleq \top
\end{align*}
where $\varphi$ is an $\FO 3$ sentence and $x$ and $y$ are arbitrary distinct variables.
Therefore the following also holds.
\begin{lemma}\label{lemma: the calculus of relations is undecidable}
  Let $A$ be any countably infinite set.
  Then both the validity problem and the finite validity problem  for $\mathrm{Eq}[\Term_{A}^{\<\cdot, \bullet^-, \cup, \bullet^{\converse}, 1\>}]$ are undecidable.
\end{lemma}

\subsection{Elimination of Relational Converse (Using the Identity Relation)}
In this subsection, we show that relational converse $\bullet^{\converse}$ can be eliminated by using fresh characters preserving validity and finite validity (Lemma~\ref{lemma: converse elimination 2}).
Let $\breve{A}$ be a countably infinite set that is disjoint from $A$.
We use $\breve{a}$ to denote the character in $\breve{A}$ denoting the converse of the character $a$ in $A$.
The following two axioms force that $\breve{a}$ is the converse of $a$:
\begin{inparaenum}[($\mathrm{C}{a}.1$)]
	\item $(a \cdot \breve{a}^{-}) \cap 1 \symboleq 0$; and
	\item $(\breve{a} \cdot a^{-}) \cap 1 \symboleq 0$.
\end{inparaenum}\\
$(\mathrm{C}{a}.\mathrm{1\mathchar`-2})$ denotes the formula $(\mathrm{C}{a}.\mathrm{1}) \symbolland (\mathrm{C}{a}.\mathrm{2})$.
In fact the following holds.
\begin{proposition}\label{proposition: converse elimination}
  For any model $\model$ over $A \cup \breve{A}$, the following hold:
  \begin{inparaenum}
    \item $\model \models (\mathrm{C}{a}.\mathrm{1\mathchar`-2}) \iff \model \models a^{\converse} \symboleq \breve{a}$;
    \item $\model \models (t \cdot u)^{\converse} \symboleq u^{\converse} \cdot t^{\converse}$; 
    \item $\model \models (t^-)^{\converse} \symboleq (t^{\converse})^-$; 
    \item $\model \models (t \cup u)^{\converse} \symboleq t^{\converse} \cup u^{\converse}$;
    \item $\model \models (t^{\converse})^{\converse} \symboleq t$;
    \item $\model \models (1)^{\converse} \symboleq 1$.
  \end{inparaenum}
\end{proposition}
\begin{proof}
  (2)--(6) are easily proved by the definition of $\jump{t}_{\model}$.
  We only prove (1).
  ($\Rightarrow$):
  We assume that there is a pair $\<v,w\>$ such that $\<v,w\> \in \jump{a}_{\model}^{\converse} \setminus \jump{\breve{a}}_{\model}$.
  Then, by $\<w,v\> \in \jump{a}_{\model}$ and $\<v,w\> \in \jump{\breve{a}^-}_{\model}$,
  $\<w,w\> \in \jump{(a \cdot \breve{a}^-) \cap 1}_{\model}$.
  However this contradicts ($\mathrm{C}{a}.1$).
  Therefore $\jump{a}_{\model}^{\converse} \subseteq \jump{\breve{a}}_{\model}$.
  We assume that there is a pair $\<v,w\>$ such that $\<v,w\> \in \jump{\breve{a}}_{\model} \setminus \jump{a}_{\model}^{\converse}$.
  Then, by $\<v,w\> \in \jump{\breve{a}}_{\model}$ and $\<w,v\> \in \jump{a^-}_{\model}$,
  $\<v,v\> \in \jump{(\breve{a} \cdot a^-) \cap 1}_{\model}$.
  However this contradicts ($\mathrm{C}{a}.2$).
  Therefore $\jump{a}_{\model}^{\converse} \supseteq \jump{\breve{a}}_{\model}$, and thus $\model \models a^{\converse} \symboleq \breve{a}$.
  ($\Leftarrow$):
  We prove the contraposition.
  If $\model \not\models (\mathrm{C}{a}.1)$, there are $v$ and $w$ such that 
    $\<v,w\> \in \jump{a}_{\model}$ and 
    $\<w,v\> \in \jump{\breve{a}^-}_{\model}$.
  Then by $\<w,v\> \in \jump{a^{\converse}}_{\model}$ and $\<w,v\> \not\in \jump{\breve{a}}_{\model}$,
  $\model \not\models a^{\converse} \symboleq \breve{a}$.
  If $\model \not\models (\mathrm{C}{a}.2)$, there are $v$ and $w$ such that 
    $\<v,w\> \in \jump{\breve{a}}_{\model}$ and 
    $\<w,v\> \in \jump{a^-}_{\model}$.
  Then by $\<v,w\> \not\in \jump{a^{\converse}}_{\model}$ and $\<v,w\> \in \jump{\breve{a}}_{\model}$,
  $\model \not\models a^{\converse} \symboleq \breve{a}$.
  \qed
\end{proof}
Let $\CF(t)$ be the normal form of a term $t$ w.r.t. the following rewriting rules:
\begin{inparaenum}
  \item $(a)^{\converse} \leadsto \breve{a}$ and $(\breve{a})^{\converse} \leadsto a$; 
  \item $(t \cdot u)^{\converse} \leadsto u^{\converse} \cdot t^{\converse}$; 
  \item $(t^-)^{\converse} \leadsto (t^{\converse})^-$; 
  \item $(t \cup u)^{\converse} \leadsto t^{\converse} \cup u^{\converse}$;
  \item $(t^{\converse})^{\converse} \leadsto t$;
  \item $(1)^{\converse} \leadsto 1$.
\end{inparaenum}
Note that, for any term $t$, $\CF(t)$ does not contain $\bullet^{\converse}$.
We use $A_{\varphi}$ to denote the finite set of all characters occurring in $\varphi$.
\begin{lemma}\label{lemma: converse elimination 2}
  For any $t, u \in \Term_{A}$, the following are equivalent:
  \begin{inparaenum}
    \item $t \symboleq u$ is valid (resp. finitely valid).
    \item $(\bigland_{a \in A_{t \symboleq u}} (\mathrm{C}{a}.\mathrm{1\mathchar`-2})) \symbolto \CF(t) \symboleq \CF(u)$ is valid (resp. finitely valid).
  \end{inparaenum}
\end{lemma}
\begin{proof}
  (2) $\Rightarrow$ (1):
  Let $\model$ be a model over $A$ such that $\model \not\models t \symboleq u$.
  We define the model $\model'$ over $A \cup \breve{A}$ by $\model' \assignment \<|\model|,\mathcal{R}'\>$, where 
  $\mathcal{R}'(a) \assignment \mathcal{R}_{\model}(a)$ for $a \in A$; and
  $\mathcal{R}'(\breve{a}) \assignment \mathcal{R}_{\model}(a)^{\converse}$ for $\breve{a} \in \breve{A}$.
  Then $\model' \models \bigland_{a \in A_{t \symboleq u}} (\mathrm{C}{a}.\mathrm{1\mathchar`-2})$ holds by Proposition~\ref{proposition: converse elimination} (1).
  Moreover, for any term $t' \in \Term_A$, $\jump{t'}_{\model} = \jump{\CF(t')}_{\model'}$ holds by using Proposition~\ref{proposition: converse elimination}.
  Therefore $\model' \not\models \CF(t) \symboleq \CF(u)$.
  (1) $\Rightarrow$ (2):
  Let $\model$ be a model over $A \cup \breve{A}$ such that $\model \models \bigland_{a \in A_{t \symboleq u}} (\mathrm{C}{a}.\mathrm{1\mathchar`-2})$ holds, but 
  $\model \not\models \CF(t) \symboleq \CF(u)$.
  Then $\jump{t'}_{\model} = \jump{\CF(t')}_{\model}$ holds for any term $t'$ over $A_{t \symboleq u}$ by using Proposition~\ref{proposition: converse elimination},
  and thus $\model \not\models t \symboleq u$.
  \qed
\end{proof}
By Lemma~\ref{lemma: equational formula}, 
each formula $(\bigland_{a \in A_{t \symboleq u}} (\mathrm{C}{a}.\mathrm{1\mathchar`-2})) \symbolto \CF(t) \symboleq \CF(u)$ can be translated to 
an equational formula without $\bullet^{\converse}$.
\subsection{Elimination of the Identity Relation}
In this subsection, we also eliminate the identity relation symbol $1$ by using a fresh character $E$.
The key is to construct an equivalence relation denoting the identity relation not using $1$ or $\bullet^{\converse}$.
Let us consider the following axioms:
\begin{inparaenum}[(${\mathrm{E}}.1$)]
	\item $E \cdot \top \symboleq \top$;
	\item $(E^- \cdot E) \cap E \symboleq 0$;
	\item $(E \cdot E^-) \cap E \symboleq 0$; and
	\item $(E \cdot E) \cap E^- \symboleq 0$.
\end{inparaenum}\\
$(\mathrm{E.1\mathchar`-4})$ denotes the formula $(\mathrm{E}.\mathrm{1}) \symbolland \dots \symbolland (\mathrm{E}.\mathrm{4})$.
\begin{proposition}\label{proposition: identity elimination}
  For any model $\model$ over $A \cup \{E\}$,
  $\model \models (\mathrm{E}.\mathrm{1\mathchar`-4})$ $\iff$
  $\jump{E}_{\model}$ is an equivalence relation on $|\model|$.
\end{proposition}
\begin{proof}
  ($\Leftarrow$): (E.1) is by the reflexivity.
  (E.4) is by the transitivity.
  (E.2) and (E.3) are by the symmetry and the transitivity.
  ($\Rightarrow$): The reflexivity is shown by (E.1) and (E.2).
  Let $v$ be any vertex in $|\model|$.
  Then let $w$ be a vertex such that $\<v,w\> \in \jump{E}_{\model}$ holds (such $w$ always exists by (E.1)).
  We assume that $\<v,v\> \not\in \jump{E}_{\model}$.
  Then, by $\<v,v\> \in \jump{E^-}_{\model}$, $\<v,w\> \in \jump{(E^- \cdot E) \cap E}_{\model}$ holds.
  However this contradicts (E.2). Therefore $\<v,v\> \in \jump{E}_{\model}$.
  The symmetry is by the reflexivity and (E.3).
  Let $\<v,w\> \in \jump{E}_{\model}$.
  We assume $\<w,v\> \not\in \jump{E}_{\model}$.
  Then $\<v,v\> \in \jump{(E \cdot E^-) \cap E}_{\model}$ by the reflexivity.
  However this contradicts (E.3). Therefore $\<w,v\> \in \jump{E}_{\model}$.
  The transitivity is by (E.4).
  \qed
\end{proof}
For every term $t$, the term $\IF(t)$ is inductively defined as follows:
\begin{inparaenum}
	\item $\IF(a) \assignment E \cdot a \cdot E$;
	\item $\IF(t \cdot u) \assignment \IF(t) \cdot \IF(u)$;
	\item $\IF(t^-) \assignment \IF(t)^-$;
	\item $\IF(t \cup u) \assignment \IF(t) \cup \IF(u)$;
	\item $\IF(t^{\converse}) \assignment \IF(t)^{\converse}$;
	\item $\IF(1) \assignment E$.
\end{inparaenum}
Note that \begin{inparaenum}[(i)]
  \item $\IF(t)$ does not contain the symbol $1$; and 
  \item if $t$ does not contain $\bullet^{\converse}$, then $\IF(t)$ also does not contain $\bullet^{\converse}$.
\end{inparaenum}
\begin{lemma}\label{lemma: identity elimination}
  For any $t, u \in \Term_{A}^{\<\cdot, \bullet^-, \cup, \bullet^{\converse}, 1\>}$,
  the following are equivalent:
  \begin{inparaenum}
    \item $t \symboleq u$ is valid (resp. finitely valid).
    \item $(\mathrm{E.1\mathchar`-4}) \symbolto \IF(t) \symboleq \IF(u)$ is valid (resp. finitely valid).
  \end{inparaenum}
\end{lemma}
\begin{proof}
  (2) $\Rightarrow$ (1):
  Let $\model$ be a model over $A$ such that $\model \not\models t \symboleq u$.
  We define the model $\model'$ over $A \cup \{E\}$ by $\model' \assignment \<|\model|,\mathcal{R}'\>$, where 
  $\mathcal{R}'(a) \assignment \mathcal{R}_{\model}(a)$ for $a \in A$ and $\mathcal{R}'(E) \assignment \triangle(|\model|)$.
  Then $\model' \models (\mathrm{E.1\mathchar`-4})$ holds because $\jump{E}_{\model'}$ is an equivalence relation.
  Also $\model' \not\models \IF(t) \symboleq \IF(u)$ holds because
  $\jump{t}_{\model} = \jump{\IF(t)}_{\model'}$ holds by the definition of $\model'$.
  (1) $\Rightarrow$ (2):
  Let $\model$ be a model over $A \cup \{E\}$ such that $\model \models (\mathrm{E.1\mathchar`-4})$ holds,
  but $\model \not\models \IF(t) \symboleq \IF(u)$.
  We define the model $\model'$ over $A$ by $\model' \assignment \<|\model|/{E},\mathcal{R}'\>$, where
  $|\model|/{E}$ is the quotient set of $|\model|$ by $\jump{E}_{\model}$;
  $[v]_E$ denotes the equivalence class of an element $v$ with respect to $\jump{E}_{\model}$; and
  $\mathcal{R}'(a) \assignment \{\<[v]_{E},[w]_{E}\> \mid \<v,w\> \in \mathcal{R}_{\model}(a)\}$ for $a \in A$.
  Then $\{\<[v]_{E},[w]_{E}\> \mid \<v,w\> \in \jump{\IF(t')}_{\model}\} = \jump{t'}_{\model'}$ holds for any term $t'$.
  This is proved by induction on the structure of $t'$ (see Appendix~\ref{appendix: identity elimination}).
  Therefore $\model' \not\models t \symboleq u$.
  \qed
\end{proof}
We can eliminate both converse and the identity relation preserving validity and finite validity by Lemmas~\ref{lemma: converse elimination 2} and \ref{lemma: identity elimination}.
Therefore the following holds.
\begin{corollary}\label{corollary: the calculus of relations is undecidable 2}
  Let $A$ be a countably infinite set.
  Then both the validity problem and the finite validity problem for $\mathrm{Eq}[\Term_{A}^{\<\cdot, \bullet^-, \cup\>}]$ are undecidable.
\end{corollary}
\section{Reductions to the One Binary Relation Case}
\label{section: Reductions to the One Binary Relation Case}
In this section we consider reducing the number of characters.
Let $A$ be an ordered set $\{a_1, a_2, \dots\}$.
Without loss of generality, we can assume that every term is in $\Term_{A}^{\<\cdot, \bullet^-, \cup\>}$ by Corollary~\ref{corollary: the calculus of relations is undecidable 2}.
In this section we give two reductions to the one binary relation case.
First we give a simple reduction, but this reduction uses the identity relation.
The second reduction is a bit more complex than the first reduction, but the reduction does not use the identity relation.

\subsection{A Conservative Reduction Using Identity}
We first give a reduction $\mathrm{T}_1$ from $\Term_{A}^{\<\cdot, \bullet^-, \cup, 1\>}$ to $\Term_{\{a\}}^{\<\cdot, \bullet^-, \cup, 1\>}$.
The term $\mathrm{T}_1(t)$ is inductively defined as follows:
\begin{inparaenum}
  \item $\mathrm{T}_1(a_i) \assignment (a \cap 1) \cdot a \cdot ((a^- \cap 1) \cdot a)^{i} \cdot (a \cap 1)$;
  \item $\mathrm{T}_1(t \cdot u) \assignment \mathrm{T}_1(t) \cdot \mathrm{T}_1(u)$;
  \item $\mathrm{T}_1(t^-) \assignment (a \cap 1) \cdot \mathrm{T}_1(t)^- \cdot (a \cap 1)$;
  \item $\mathrm{T}_1(t \cup u) \assignment \mathrm{T}_1(t) \cup \mathrm{T}_1(u)$;
  \item $\mathrm{T}_1(1) \assignment (a \cap 1)$.
\end{inparaenum}

\begin{figure}[htbp]
	\centering
	\includegraphics{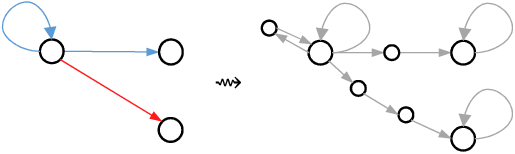}
	\caption{A transformation to the one binary relation case using identity}
	\label{figure: to_One}
\end{figure}

Figure~\ref{figure: to_One} is an example of transforming models over $A$ to models over $\{a\}$ in Lemma~\ref{lemma: to one using identity}.
Each blue (resp. red, gray) colored edge denotes an edge labeled with $a_1$ (resp. $a_2$, $a$).
\begin{lemma}\label{lemma: to one using identity}
  For any $t, u \in \Term_{A}^{\<\cdot, \bullet^-, \cup, 1\>}$,
  the following are equivalent:
  \begin{inparaenum}
    \item $t \symboleq u$ is valid (resp. finitely valid).
    \item $\mathrm{T}_1(t) \symboleq \mathrm{T}_1(u)$ is valid (resp. finitely valid).
  \end{inparaenum}
\end{lemma}
\begin{proof}
  (2) $\Rightarrow$ (1):
  Let $\model$ be a model over $A_{t \symboleq u}$ such that $\model \not\models t \symboleq u$.
  We define the model $\model'$ over $\{a\}$ by $\model' \assignment \<|\model'|,\mathcal{R}'\>$, where
  $|\model'| \assignment |\model| \cup \{\<v,w,i,j\> \mid \<v,w\> \in \mathcal{R}_{\model}(a_i), j \in [i] \}$ and
  $\mathcal{R}'(a) \assignment \triangle(|\model|) 
    \cup \{\<v,\<v,w,i,1\>\>,\<\<v,w,i,i\>,w\> \mid \<v,w\> \in \mathcal{R}_{\model}(a_i)\}
    \cup \{\<\<v,w,i,j\>,\<v,w,i,j+1\>\> \mid \<v,w\> \in \mathcal{R}_{\model}(a_i), j \in [i-1] \}$.
  The right-hand side in Figure~\ref{figure: to_One} is an example of $\model'$.
  Note that if $\model$ is finite, $\model'$ is also finite because $A_{t \symboleq u}$ is finite.
  Then $\jump{t}_{\model} = \jump{\mathrm{T}_1(t)}_{\model'}$ holds.
  This is easily proved by induction on the structure of $t$ (see Appendix~\ref{appendix: to one using identity 1}).
  Therefore $\model' \not\models \mathrm{T}_1(t) \symboleq \mathrm{T}_1(u)$.
  (1) $\Rightarrow$ (2):
  Let $\model$ be a model over $\{a\}$ such that $\model \not\models \mathrm{T}_1(t) \symboleq \mathrm{T}_1(u)$.
  Then we define the model $\model'$ over $A$ by $\model' \assignment \<|\model'|, \mathcal{R}'\>$, where
  $|\model'| \assignment \{v \mid \<v,v\> \in \jump{a}_{\model}\}$ and
  $\mathcal{R}'(a_i) \assignment |\model'|^2 \cap \jump{\mathrm{T}_1(a_i)}_{\model}$.
  Here $|\model'| \neq \emptyset$, since if not, $\jump{\mathrm{T}_1(t)}_{\model} = \emptyset = \jump{\mathrm{T}_1(u)}_{\model}$.
  Then $\jump{t}_{\model'} = \jump{\mathrm{T}_1(t)}_{\model}$ holds.
  This is easily proved by induction on the structure of $t$ (see Appendix~\ref{appendix: to one using identity 2}).
  Therefore $\model' \not\models t \symboleq u$. \qed
\end{proof}
By the above lemma, the following has been proved.
\begin{theorem}\label{theorem: one variable rel undecidable}
  Let $a$ be a character.
  Then both the validity problem and the finite validity problem  for $\mathrm{Eq}[\Term_{\{a\}}^{\<\cdot, \bullet^-, \cup, 1\>}]$ are undecidable.
\end{theorem}
\begin{remark}
  The undecidability of the validity problem for the calculus of relations with just one character over the signature $\<\cdot, \bullet^{-}, \cup, \bullet^{\converse}, 1\>$ had been shown by Maddux \cite[p.~399]{madduxUndecidableSemiassociativeRelation1994}.
  More strongly, Theorem~\ref{theorem: one variable rel undecidable} shows that
  the validity problem is undecidable even without $\bullet^{\converse}$
  and shows that the finite validity problem is also undecidable.
  Furthermore, Theorem~\ref{theorem: one variable rel undecidable} will be strengthened to Theorem~\ref{theorem: to one with none}.
\end{remark}
Combining Theorem~\ref{theorem: one variable rel undecidable} and Lemma~\ref{lemma: identity elimination}, the following is also proved.
\begin{corollary}\label{corollary: two variable rel undecidable}
  Let $a_1, a_2$ be two distinct characters.
  Then both the validity problem and the finite validity problem for $\mathrm{Eq}[\Term_{\{a_1,a_2\}}^{\<\cdot, \bullet^-, \cup\>}]$ are undecidable.
\end{corollary}

\subsection{A Conservative Reduction Not Using Identity}
  In this subsection we give another reduction, which does not use the identity relation or relational converse.
  The key is how to distinguish the vertices of a given model from the other vertices on the transformed model, not relying on the identity relation.
  Without loss of generality we can assume that the number of characters is $2$ by Corollary~\ref{corollary: two variable rel undecidable}.
  Now we consider the following axioms:
  \begin{inparaenum}[($\mathrm{Ax}.1$)]
    \item $(a \cap a^2) \cdot ((a \cap a^3) \cdot \top)^- \symboleq 0$; and
    \item $(\top \cdot (a \cap a^2))^- \cdot (a \cap a^3) \symboleq 0$.
  \end{inparaenum}
  $(\mathrm{Ax.1\mathchar`-2})$ denotes the formula $(\mathrm{Ax.1}) \symbolland (\mathrm{Ax.2})$.
  These axioms are used to force $\cod(\jump{a \cap a^2}_{\model}) = \dom(\jump{a \cap a^3}_{\model})$,
  where $\dom(R)$ (resp. $\cod(R)$) denotes the \emph{domain} (resp. \emph{codomain}) of a binary relation $R$,
  i.e., $\dom(R) \assignment \{v \mid \exists w.\<v,w\> \in R\}$ and $\cod(R) \assignment \{w \mid \exists v.\<v,w\> \in R\}$.
  In fact the following holds.
  \begin{proposition}\label{proposition: dom and cod}
    For every model $\model$ over $\{a\}$,
    $\model \models (\mathrm{Ax.1\mathchar`-2})$
    $\iff$
    $\cod(\jump{a \cap a^2}_{\model}) = \dom(\jump{a \cap a^3}_{\model})$.
  \end{proposition}
  \begin{proof}
    By 
    $\model \models (\mathrm{Ax.1}) \iff \cod(\jump{a \cap a^2}_{\model}) \subseteq \dom(\jump{a \cap a^3}_{\model})$ and
    $\model \models (\mathrm{Ax.2}) \iff \cod(\jump{a \cap a^2}_{\model}) \supseteq \dom(\jump{a \cap a^3}_{\model})$.
    \qed
  \end{proof}
  We define a reduction $\mathrm{T}_2$ from $\Term_{\{a_1,a_2\}}^{\<\cdot, \bullet^-, \cup\>}$ to $\Term_{\{a\}}^{\<\cdot, \bullet^-, \cup\>}$ as follows:
  \begin{inparaenum}
    \item $\mathrm{T}_2(a_i) \assignment (a \cap a^3) \cdot a^{i+1} \cdot (a \cap a^2)$;
    \item $\mathrm{T}_2(t \cdot u) \assignment \mathrm{T}_2(t) \cdot \mathrm{T}_2(u)$;
    \item $\mathrm{T}_2(t^-) \assignment ((a \cap a^3) \cdot \top) \cap (\top \cdot (a \cap a^2)) \cap \mathrm{T}_2(t)^-$;
    \item $\mathrm{T}_2(t \cup u) \assignment \mathrm{T}_2(t) \cup \mathrm{T}_2(u)$.
  \end{inparaenum}
  Intuitively, both the codomain of $a \cap a^2$ and the domain of $a \cap a^3$ denote the set of vertices in the pre-transformed model in the above reduction.

\begin{figure}[htbp]
    \centering
    \includegraphics{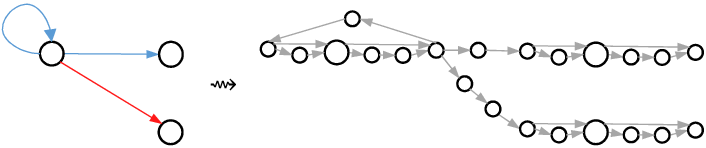}
    \caption{A transformation to the one binary relation case not using identity}
    \label{figure: to_One_with_none}
  \end{figure}

  Figure~\ref{figure: to_One_with_none} is an example of transforming models over $\{a_1,a_2\}$ to models over $\{a\}$ in Lemma~\ref{lemma: to one with none}.
  Each blue (resp. red, gray) colored edge denotes an edge labeled with $a_1$ (resp. $a_2$, $a$).
  \vspace{-1em}
  \begin{lemma}\label{lemma: to one with none}
    For any $t, u \in \Term_{\{a_1,a_2\}}^{\<\cdot, \bullet^-, \cup\>}$,
    the following are equivalent:
    \begin{inparaenum}
      \item $t \symboleq u$ is valid (resp. finitely valid).
      \item $(\mathrm{Ax.1\mathchar`-2}) \symbolto \mathrm{T}_2(t) \symboleq \mathrm{T}_2(u)$ is valid (resp. finitely valid).
    \end{inparaenum}
  \end{lemma}
  \begin{proof}
    (2) $\Rightarrow$ (1):
    Let $\model$ be a model over $\{a_1,a_2\}$ such that $\model \not\models t \symboleq u$.
    We define the model $\model'$ over $\{a\}$ by $\model' \assignment \<|\model'|,\mathcal{R}'\>$, where
    $|\model'| \assignment (|\model| \times [6]) \cup \{\<v,w,i,j\> \mid \<v,w\> \in \mathcal{R}_{\model}(a_i), j \in [i] \}$ and
    $\mathcal{R}'(a)$ is the union of the following: 
    \begin{enumerate}[(i)]
      \item $\bigcup_{v \in |\model|} (\{\<\<v,i\>,\<v,i+1\>\> \mid i \in [5]\} \cup \{\<\<v,1\>,\<v,3\>\>,\<\<v,3\>,\<v,6\>\>\})$;
      \item $\bigcup_{i \in \{1,2\}} \{\<\<v,6\>,\<v,w,i,1\>\>,\<\<v,w,i,i\>,\<w,1\>\> \mid \<v,w\> \in \mathcal{R}_{\model}(a_i)\}$; and
      \item $\{\<\<v,w,2,1\>,\<v,w,2,2\>\> \mid \<v,w\> \in \mathcal{R}_{\model}(a_2)\}$.
    \end{enumerate}
    The right-hand side in Figure~\ref{figure: to_One_with_none} is an example of $\model'$.
    Note that if $\model$ is finite, $\model'$ is also finite.

    First $\model' \models (\mathrm{Ax.1\mathchar`-2})$ holds by
    $\cod(\jump{a \cap a^2}_{\model'}) = \{\<v,3\> \mid v \in |\model|\} = \dom(\jump{a \cap a^3}_{\model'})$. 
    Moreover the following holds: $\{\<\<v,3\>, \<w,3\>\> \mid \<v,w\> \in \jump{t}_{\model} \} = \jump{\mathrm{T}_2(t)}_{\model'}$ $\cdots$ ($\heartsuit$).
    $\model' \not\models \mathrm{T}_2(t) \symboleq \mathrm{T}_2(u)$ is proved by ($\heartsuit$).
    We prove ($\heartsuit$) by induction on the structure of $t$.\\
    If $t \equiv a_i$, let $\<v,w\>$ be a pair such that $\<v,w\> \in \jump{a_i}_{\model}$.
    Then $\<\<v,3\>,\<w,3\>\> \in \jump{(a \cap a^3) \cdot a^{i+1} \cdot (a \cap a^2)}_{\model'}$ is easily checked.
    Conversely, if $\<v,w\> \in \jump{(a \cap a^3) \cdot a^{i+1} \cdot (a \cap a^2)}_{\model'}$, 
    then there are $v',w'\in |\model|$ such that $\<v,w\> = \<\<v',3\>,\<w',3\>\>$ 
    by $\cod(\jump{a \cap a^2}_{\model'}) = \dom(\jump{a \cap a^3}_{\model'}) = \{\<v,3\> \mid v \in |\model|\}$.
    Then $\{v \mid \<\<v',3\>, v\> \in \jump{a \cap a^3}_{\model'}\} = \{\<v',6\>\}$ and 
    $\{w \mid \<w,\<w',3\>\> \in \jump{a \cap a^2}_{\model'}\} = \{\<w',1\>\}$ hold, and thus
    $\<\<v',6\>,\<w',1\>\> \in \jump{a^{i+1}}_{\model'}$.
    If we assume $\<v',w'\> \not\in \jump{a_i}_{\model}$, then 
    $\<w',1\>$ is not reachable from $\<v',6\>$ in $i + 1$ steps because
    $i + 1 \le 3$, the length of paths from $\<v',6\>$ to $\<w',1\>$ via $\<v',w',j,1\>$ with $j \neq i$ is $j+1$,
    and the length of paths from $\<v',6\>$ to $\<w',1\>$ via some $\<w'',1\>$ is at least $6$,
    where $w''$ is an element in $|\model|$ not equal to $w'$.
    (More concretely, the following path is the shortest path from $\<v',6\>$ to $\<w',1\>$ via $\<w'',1\>$:
    $\<v',6\> \leadsto \<v',w'',1, 1\> \leadsto \<w'',1\> \leadsto \<w'',3\> \leadsto \<w'',6\> \leadsto \<w'',w',1, 1\> \leadsto \<w',1\>$.)
    However this contradicts $\<\<v',6\>,\<w',1\>\> \in \jump{a^{i+1}}_{\model'}$.
    Therefore $\<v',w'\> \in \jump{a_i}_{\model}$.\\
    If $t \equiv t \cdot u$:
    {\small
    \begin{align*}
      &\jump{\mathrm{T}_2(t \cdot u)}_{\model'}
      = \jump{\mathrm{T}_2(t) \cdot \mathrm{T}_2(u)}_{\model'}
      = \jump{\mathrm{T}_2(t)}_{\model'} \relcdot \jump{\mathrm{T}_2(u)}_{\model'}\\
      &= \{\<\<v,3\>,\<v'',3\>\> \mid \<v,v''\> \in \jump{t}_{\model} \} 
      \relcdot \{\<\<v'',3\>,\<v',3\>\> \mid \<v'',v'\> \in \jump{u}_{\model} \} \tag{I.H.}\\
      &= \{\<\<v,3\>,\<v',3\>\> \mid \<v,v'\> \in \jump{t \cdot u}_{\model} \}.
    \end{align*}
    }
    If $t \equiv t^-$:
    {\small
    \begin{align*}
      \jump{\mathrm{T}_2(t^-)}_{\model'}
      &= \jump{((a \cap a^3) \cdot \top) \cap (\top \cdot (a \cap a^2)) \cap \mathrm{T}_2(t)^-}_{\model'}\\
      &= \{\<\<v,3\>,\<w,3\>\> \mid \<v,w\> \in |\model|^2\} \cap \jump{\mathrm{T}_2(t)^-}_{\model'}\\
      &= \{\<\<v,3\>,\<w,3\>\> \mid \<v,w\> \in |\model|^2 \} \setminus \jump{\mathrm{T}_2(t)}_{\model'}\\
      &= \{\<\<v,3\>,\<w,3\>\> \mid \<v,w\> \in |\model|^2 \} \setminus \{\<\<v,3\>,\<w,3\>\> \mid \<v,w\> \in \jump{t}_{\model}\} \tag{I.H.}\\
      &= \{\<\<v,3\>,\<w,3\>\> \mid \<v,w\> \in \jump{t^-}_{\model}\}.
    \end{align*}
    }
    If $t \equiv t \cup u$:
    {\small
    \begin{align*}
      &\jump{\mathrm{T}_2(t \cup u)}_{\model'} = \jump{\mathrm{T}_2(t) \cup \mathrm{T}_2(u)}_{\model'} = \jump{\mathrm{T}_2(t)}_{\model'} \cup \jump{\mathrm{T}_2(u)}_{\model'}\\
      &= \{\<\<v,3\>,\<w,3\>\> \mid \<v,w\> \in \jump{t}_{\model} \} 
      \cup \{\<\<v,3\>,\<w,3\>\> \mid \<v,w\> \in \jump{u}_{\model} \} \tag{I.H.}\\
      &= \{\<\<v,3\>,\<w,3\>\> \mid \<v,w\> \in \jump{t \cup u}_{\model} \}.
    \end{align*}
    }
    \vspace{-1em}

    (1) $\Rightarrow$ (2):
    Let $\model$ be a model over $\{a\}$ such that 
    $\model \models (\mathrm{Ax.1\mathchar`-2})$, but 
    $\model \not\models \mathrm{T}_2(t) \symboleq \mathrm{T}_2(u)$.
    Then we define the model $\model'$ over $\{a_1,a_2\}$ by $\model' \assignment \<|\model'|, \mathcal{R}'\>$, where
    $|\model'| \assignment \dom(\jump{a \cap a^3}_{\model})$ and
    $\mathcal{R}'(a_i) \assignment \jump{\mathrm{T}_2(a_i)}_{\model}$.
    Note that $\mathcal{R}'(a_i) \subseteq |\model'|^2$ holds because 
    $\dom(\jump{a \cap a^3}_{\model}) = \cod(\jump{a \cap a^2}_{\model})$ by Proposition~\ref{proposition: dom and cod}.
    Here $|\model'| \neq \emptyset$, since if not, $\jump{\mathrm{T}_2(t)}_{\model} = \emptyset = \jump{\mathrm{T}_2(u)}_{\model}$.
    Then the following holds: $\jump{t}_{\model'} = \jump{\mathrm{T}_2(t)}_{\model}$ $\cdots$ ($\spadesuit$).
    $\model' \not\models t \symboleq u$ is proved by ($\spadesuit$).
    We now prove ($\spadesuit$) by induction on the structure of $t$.\\
    If $t \equiv a_i$, $\jump{a_i}_{\model'} = \jump{\mathrm{T}_2(a_i)}_{\model}$ is shown by the definition of $\model'$.\\
    If $t \equiv t \cdot u$,
    \vspace{-1em}
    {\small
    \begin{align*}
      \jump{t \cdot u}_{\model'}
      &= \jump{t}_{\model'} \relcdot \jump{u}_{\model'}
      = \jump{\mathrm{T}_2(t)}_{\model} \relcdot \jump{\mathrm{T}_2(u)}_{\model} \tag{I.H.}\\
      &= \jump{\mathrm{T}_2(t) \cdot \mathrm{T}_2(u)}_{\model}
      = \jump{\mathrm{T}_2(t \cdot u)}_{\model}.
    \end{align*}
    }
    If $t \equiv t^-$,
    \vspace{-1em}
    {\small
    \begin{align*}
      \jump{t^-}_{\model'}
      &= |\model'|^2 \setminus \jump{t}_{\model'}
      = |\model'|^2 \setminus \jump{\mathrm{T}_2(t)}_{\model} \tag{I.H.}\\
      &= \jump{((a \cap a^3) \cdot \top) \cap (\top \cdot (a \cap a^2))}_{\model} \setminus \jump{\mathrm{T}_2(t)}_{\model}\\
      &= \jump{((a \cap a^3) \cdot \top) \cap (\top \cdot (a \cap a^2)) \cap \mathrm{T}_2(t)^-}_{\model}
      = \jump{\mathrm{T}_2(t^-)}_{\model}.
    \end{align*}
    }
    If $t \equiv t \cup u$,
    \vspace{-1em}
    {\small
    \begin{align*}
      &\jump{t \cup u}_{\model'}
      = \jump{t}_{\model'} \cup \jump{u}_{\model'}
      = \jump{\mathrm{T}_2(t)}_{\model} \cup \jump{\mathrm{T}_2(u)}_{\model} \tag{I.H.}\\
      &= \jump{\mathrm{T}_2(t) \cup \mathrm{T}_2(u)}_{\model}
      = \jump{\mathrm{T}_2(t \cup u)}_{\model}.
    \end{align*}
    }
    \vspace{-2em}
    \qed
  \end{proof}
  By Lemma~\ref{lemma: to one with none}, the next theorem has been proved.
  \begin{theorem}\label{theorem: to one with none}
    Let $a$ be a character.
    Then both the validity problem and the finite validity problem for $\mathrm{Eq}[\Term_{\{a\}}^{\<\cdot, \bullet^-, \cup\>}]$ are undecidable.
  \end{theorem}
  Combining Theorem~\ref{theorem: to one with none} and Lemma~\ref{lemma: rel to FO3}, the following is also proved.
  Note that, if a term $t$ does not contain $1$, the formula $G(t,i,j)$ does not contain $=$.
  \begin{theorem}\label{theorem: Church-Herbrand FO3}
    Let $a$ be a character.
    Then both the validity problem and the finite validity problem for $\FO 3_{\{a\}}$ are undecidable.
  \end{theorem}

  \begin{remark*}[Correction of \cite{nakamuraUndecidabilityFO3Calculus2019}]
  In the conference version \cite{nakamuraUndecidabilityFO3Calculus2019},
  the axioms (Ax.1) and (Ax.2) were given as
  $(a \cap a^2) \cdot \top \symboleq (a \cap a^2) \cdot (a \cap a^3) \cdot \top$ and
  $\top \cdot (a \cap a^3) \symboleq \top \cdot (a \cap a^2) \cdot (a \cap a^3)$.
  However, Proposition~\ref{proposition: dom and cod} fails for them.
  Proposition~\ref{proposition: dom and cod} holds for the corrected axioms (Ax.1) and (Ax.2) above.
  The rest of the proof is unchanged, as the axioms (Ax.1) and (Ax.2) are used only through Proposition~\ref{proposition: dom and cod}.
  \end{remark*}

\section{Conclusion}\label{section: conclusion}
In this paper we showed that the validity problem and the finite validity problem are undecidable for the following classes:
\begin{inparaenum}
  \item $\FO 3$ with just one binary relation symbol and without equality; and 
  \item equational formulas of the calculus of relations with just one character over the signature $\<\cdot,\bullet^-,\cup\>$.
\end{inparaenum}
In connection with (2), the following decidable fragments are known.
\begin{theorem}[cf.\ {\cite[Theorem 5]{andrekaEquationalTheoryUnionfree1995}} for (1)]\label{theorem: decidable subclasses}
  The validity problem and the finite validity problem are decidable for equational formulas of the following classes:
  \begin{inparaenum}
		\item the calculus of relations over the signature $\<\cdot,\cup\>$ (even with $\bullet^\converse$, $1$, $0$, $\top$, $\cap$, and atomic negation); and
    \item the calculus of relations over the signature $\<\bullet^-,\cup\>$ (even with $\bullet^\converse$, $1$).
	\end{inparaenum}
\end{theorem}
These decidability results are proved by the reduction to $\FO 3$ (Lemma~\ref{lemma: rel to FO3}) and using the decidability of the satisfiability problem and the finite model property of the $\symbolexists^* \symbolforall^*$ case with equality (proved by Ramsey \cite{ramseyProblemFormalLogic1930} in 1930, see also \cite[Sect. 6.2.2]{borgerClassicalDecisionProblem1997}).
To the best of our knowledge, the (un)decidability of the validity problem and the finite validity problem for equational formulas of the calculus of relations over the signature $\<\cdot, \bullet^{-}\>$ is open regardless of the number of characters.

\bibliographystyle{splncs04}
\bibliography{main}

\appendix
  \section{Completion of the Proof of Lemma~\ref{lemma: identity elimination}}\label{appendix: identity elimination}
    We prove the following by induction on the structure of $t$:
    \begin{enumerate}
      \item If $\<v,v'\>, \<w,w'\> \in \jump{E}_{\model}$,
          then $\<v,w\> \in \jump{\IF(t)}_{\model} \iff \<v',w'\> \in \jump{\IF(t)}_{\model}$; and
      \item $\{\<[v]_{E},[w]_{E}\> \mid \<v,w\> \in \jump{\IF(t)}_{\model}\} = \jump{t}_{\model'}$,
    \end{enumerate}
    where $\model$ and $\model'$ are the models in Lemma~\ref{lemma: identity elimination}.
    For (1), if $t \equiv a$ or $t \equiv 1$, this part is proved by the definition of $\IF(t)$ and that $\jump{E}_{\model}$ is an equivalence relation.
    If $t \equiv t \cdot u$, 
    \begin{align*}
      \<v,w\> \in \jump{\IF(t \cdot u)}_{\model}
      &\iff \exists x. \<v,x\> \in \jump{\IF(t)}_{\model} \land \<x,w\> \in \jump{\IF(u)}_{\model}\\
      &\iff \exists x. \<v',x\> \in \jump{\IF(t)}_{\model} \land \<x,w'\> \in \jump{\IF(u)}_{\model} \tag{I.H.}\\
      &\iff \<v',w'\> \in \jump{\IF(t \cdot u)}_{\model}.
    \end{align*}
    If $t \equiv t^-$,
    \begin{align*}
      \<v,w\> \in \jump{\IF(t^-)}_{\model}
      \iff \<v,w\> \not\in \jump{\IF(t)}_{\model}
      &\iff \<v',w'\> \not\in \jump{\IF(t)}_{\model} \tag{I.H.}\\
      &\iff \<v',w'\> \in \jump{\IF(t^-)}_{\model}.
    \end{align*}
    If $t \equiv t \cup u$, 
    \begin{align*}
      \<v,w\> \in \jump{\IF(t \cup u)}_{\model}
      &\iff \<v,w\> \in \jump{\IF(t)}_{\model} \text{ or } \<v,w\> \in \jump{\IF(u)}_{\model}\\
      &\iff \<v',w'\> \in \jump{\IF(t)}_{\model} \text{ or } \<v',w'\> \in \jump{\IF(u)}_{\model} \tag{I.H.}\\
      &\iff \<v',w'\> \in \jump{\IF(t \cup u)}_{\model}.
    \end{align*}
    If $t \equiv t^{\converse}$, 
    \begin{align*}
      \<v,w\> \in \jump{\IF(t^{\converse})}_{\model}
      \iff \<w,v\> \in \jump{\IF(t)}_{\model}
      &\iff \<w',v'\> \in \jump{\IF(t)}_{\model} \tag{I.H.}\\
      &\iff \<v',w'\> \in \jump{\IF(t^{\converse})}_{\model}.
    \end{align*}

    For (2), if $t \equiv a$ or $t \equiv 1$, this part is proved by the definition of $\model'$.
    If $t \equiv t \cdot u$,
    {\small
    \begin{align*}
      \jump{t \cdot u}_{\model'}
      &= \jump{t}_{\model'} \relcdot \jump{u}_{\model'}\\
      &= \{\<[v]_{E},[v'']_{E}\> \mid \<v,v''\> \in \jump{\IF(t)}_{\model}\} \relcdot \{\<[v'']_{E},[v']_{E}\> \mid \<v'',v'\> \in \jump{\IF(u)}_{\model}\} \tag{I.H.}\\
      &= \{\<[v]_{E},[v']_{E}\> \mid \<v,v'\> \in \jump{\IF(t)}_{\model} \relcdot \jump{\IF(u)}_{\model}\} \tag{By (1)}\\
      &= \{\<[v]_{E},[v']_{E}\> \mid \<v,v'\> \in \jump{\IF(t) \cdot \IF(u)}_{\model}\}\\
      &= \{\<[v]_{E},[v']_{E}\> \mid \<v,v'\> \in \jump{\IF(t \cdot u)}_{\model}\}.
    \end{align*}
    }
    If $t \equiv t^-$,
    \begin{align*}
      \jump{t^-}_{\model'}
      &= |\model'|^2 \setminus \jump{t}_{\model'}\\
      &= \{\<[v]_{E},[v']_{E}\> \mid \<v,v'\> \in |\model|^2\} \setminus \{\<[v]_{E},[v']_{E}\> \mid \<v,v'\> \in \jump{\IF(t)}_{\model}\} \tag{I.H.}\\
      &= \{\<[v]_{E},[v']_{E}\> \mid \<v,v'\> \in |\model|^2 \setminus \jump{\IF(t)}_{\model}\} \tag{By (1)}\\
      &= \{\<[v]_{E},[v']_{E}\> \mid \<v,v'\> \in \jump{\IF(t^-)}_{\model}\}.
    \end{align*}
    
    If $t \equiv t \cup u$,
    {\small
    \begin{align*}
      \jump{t \cup u}_{\model'}
      &= \jump{t}_{\model'} \cup \jump{u}_{\model'}\\
      &= \{\<[v]_{E},[v']_{E}\> \mid \<v,v'\> \in \jump{\IF(t)}_{\model}\} \cup \{\<[v]_{E},[v']_{E}\> \mid \<v,v'\> \in \jump{\IF(u)}_{\model}\} \tag{I.H.}\\
      &= \{\<[v]_{E},[v']_{E}\> \mid \<v,v'\> \in \jump{\IF(t)}_{\model} \cup \jump{\IF(u)}_{\model}\}\\
      &= \{\<[v]_{E},[v']_{E}\> \mid \<v,v'\> \in \jump{\IF(t) \cup \IF(u)}_{\model}\}\\
      &= \{\<[v]_{E},[v']_{E}\> \mid \<v,v'\> \in \jump{\IF(t \cup u)}_{\model}\}.
    \end{align*}
    }
    If $t \equiv t^{\converse}$,
    \begin{align*}
      \jump{t^{\converse}}_{\model'}
      &= \jump{t}_{\model'}^{\converse} = \{\<[v]_{E},[v']_{E}\> \mid \<v,v'\> \in \jump{\IF(t)}_{\model}\}^{\converse} \tag{I.H.}\\
      &= \{\<[v]_{E},[v']_{E}\> \mid \<v,v'\> \in \jump{\IF(t)^{\converse}}_{\model}\}\\
      &= \{\<[v]_{E},[v']_{E}\> \mid \<v,v'\> \in \jump{\IF(t^{\converse})}_{\model}\}.
    \end{align*}

  \section{Completion of the Proof of Lemma~\ref{lemma: to one using identity} (2) $\Rightarrow$ (1)}\label{appendix: to one using identity 1}
    We prove $\jump{t}_{\model} = \jump{\mathrm{T}_1(t)}_{\model'}$
    by induction on the structure of $t$, where $\model$ and $\model'$ are the models in Lemma~\ref{lemma: to one using identity}.\\
    If $t \equiv a_i$,
    let $\<v,w\>$ be a pair such that $\<v,w\> \in \jump{a_i}_{\model}$.
    Then $\<v,w\> \in \jump{\mathrm{T}_1(a_i)}_{\model'}$ holds by the construction of $\model'$.
    Conversely if $\<v,w\> \in \jump{\mathrm{T}_1(a_i)}_{\model'}$,
    there is a sequence $\<v_0,\dots,v_{i+1}\>$ such that 
    \begin{inparaenum}
      \item $\<v,w\> = \<v_0,v_{i+1}\>$;
      \item $v_0, v_{i+1} \in \dom(\jump{a \cap 1}_{\model'})$ and $v_1, \dots, v_i \not\in \dom(\jump{a \cap 1}_{\model'})$; and
      \item $\<v_j,v_{j+1}\> \in \jump{a}_{\model'}$ for $0 \le j \le i$.
    \end{inparaenum}
    Therefore, $v, w \in |\model|$, and thus $\<v,w\> \in \jump{a_i}_{\model}$.\\
    If $t \equiv t \cdot u$,
    \begin{align*}
      \jump{\mathrm{T}_1(t \cdot u)}_{\model'}
      = \jump{\mathrm{T}_1(t)}_{\model'} \relcdot \jump{\mathrm{T}_1(u)}_{\model'}
      = \jump{t}_{\model} \relcdot \jump{u}_{\model}
      = \jump{t \cdot u}_{\model}. \tag{I.H.}
    \end{align*}
    If $t \equiv t^-$,
    \begin{align*}
      \jump{\mathrm{T}_1(t^-)}_{\model'}
      &= \jump{(a \cap 1) \cdot \mathrm{T}_1(t)^- \cdot (a \cap 1)}_{\model'}
      = |\model|^2 \cap (|\model'|^2 \setminus \jump{\mathrm{T}_1(t)}_{\model'})\\
      &= |\model|^2 \setminus \jump{\mathrm{T}_1(t)}_{\model'}
      = |\model|^2 \setminus \jump{t}_{\model}
      = \jump{t^-}_{\model}. \tag{I.H.}
    \end{align*}
    If $t \equiv t \cup u$,
    \begin{align*}
      \jump{\mathrm{T}_1(t \cup u)}_{\model'}
      &= \jump{\mathrm{T}_1(t)}_{\model'} \cup \jump{\mathrm{T}_1(u)}_{\model'}
      = \jump{t}_{\model} \cup \jump{u}_{\model}
      = \jump{t \cup u}_{\model}. \tag{I.H.}
    \end{align*}
    If $t \equiv 1$, by the construction of $\model'$,
    $\jump{a \cap 1}_{\model'} = \jump{1}_{\model}$, and thus
    $\jump{\mathrm{T}_1(1)}_{\model'} = \jump{1}_{\model}$.
\section{Completion of the Proof of Lemma~\ref{lemma: to one using identity} (1) $\Rightarrow$ (2)}\label{appendix: to one using identity 2}
    We prove $\jump{t}_{\model'} = \jump{\mathrm{T}_1(t)}_{\model}$
    by induction on the structure of $t$, where $\model$ and $\model'$ are the models in Lemma~\ref{lemma: to one using identity}.\\
    If $t \equiv a_i$,
    by the construction of $\model'$, $\jump{a_i}_{\model'} = \jump{\mathrm{T}_1(a_i)}_{\model}$ holds.\\
    If $t \equiv t \cdot u$,
    \begin{align*}
      \jump{\mathrm{T}_1(t \cdot u)}_{\model}
      = \jump{\mathrm{T}_1(t)}_{\model} \relcdot \jump{\mathrm{T}_1(u)}_{\model}
      = \jump{t}_{\model'} \relcdot \jump{u}_{\model'}
      = \jump{t \cdot u}_{\model'}. \tag{I.H.}
    \end{align*}
    If $t \equiv t^-$,
    \begin{align*}
      \jump{\mathrm{T}_1(t^-)}_{\model}
      &= \jump{(a \cap 1) \cdot \mathrm{T}_1(t)^- \cdot (a \cap 1)}_{\model}
      = |\model'|^2 \cap (|\model|^2 \setminus \jump{\mathrm{T}_1(t)}_{\model})\\
      &= |\model'|^2 \setminus \jump{\mathrm{T}_1(t)}_{\model}
      = |\model'|^2 \setminus \jump{t}_{\model'}
      = \jump{t^-}_{\model'}. \tag{I.H.}
    \end{align*}
    If $t \equiv t \cup u$,
    \begin{align*}
      \jump{\mathrm{T}_1(t \cup u)}_{\model}
      &= \jump{\mathrm{T}_1(t)}_{\model} \cup \jump{\mathrm{T}_1(u)}_{\model}
      = \jump{t}_{\model'} \cup \jump{u}_{\model'}
      = \jump{t \cup u}_{\model'}. \tag{I.H.}
    \end{align*}
    If $t \equiv 1$, by the construction of $\model'$,
    $\jump{1}_{\model'} = \jump{a \cap 1}_{\model}$, and thus
    $\jump{1}_{\model'} = \jump{\mathrm{T}_1(1)}_{\model}$.
\end{document}